# Limitations of Proprioceptive Working Memory

Caitlin Callaghan
Department of Mechanical and Aerospace Engineering
*University of California Irvine*
Irvine, CA United States
callaghc@uci.edu

David J. Reinkensmeyer
Department of Mechanical and Aerospace Engineering
Department of Anatomy and Neurobiology
*University of California Irvine*
Irvine, CA United States
dreinken@uci.edu

*Abstract*— Recalling previously experienced movements is essential for a range of activities, including sports, music, and rehabilitation, yet little is known about the accuracy and decay of proprioceptive working memory. We examined how introducing a short-term memory component affected movement reproduction accuracy by comparing movement reproduction under two conditions: simultaneous reproduction (SimRep) and memorized reproduction (MemRep). In Experiment 1 (N = 191), participants felt a 5-s haptic trajectory with one hand and reproduced it with the other hand simultaneously or immediately after the template ended. Errors were greater in MemRep than SimRep (31.1° vs. 21.5°, p < 0.001). MemRep trajectories showed systematic temporal distortions: participants lagged fast movements and led slow ones (R = –0.32, p = 0.01), unlike the ~270 ms lag in SimRep. In Experiment 2 (N = 33), we varied template durations (2–8 s). Longer durations increased error for MemRep but not SimRep (p < 0.001). During MemRep, accuracy declined steadily, with replay–template correlations dropping from ~0.4 to ~0.1 over ~3 s, while SimRep correlations rose from ~0.25 to ~0.6. In ~10% of MemRep templates, participants moved in the wrong direction initially, especially for low-amplitude movements (p < 0.001). Templates with more than four movements showed element omission; after four movements had been reproduced participants ceased movement prematurely, affecting up to 40% of 8-s templates. These findings show that transferring proprioceptive experiences into working memory introduces systematic temporal and structural distortions. Accuracy decays within seconds, and memory span for movement trajectories was limited to four movements.*

## I. INTRODUCTION

When developing new motor skills or recovering motor skills following injury, a common approach is for an instructor to passively guide the learner through the ideal form. The learner understands this movement through their proprioception, their sense of limb position and movement through space. Such guided movement training relies on accurate encoding and recall of the proprioceptively demonstrated movements. Then, when a trainee attempts to reproduce the movement, they must compare their attempt with the memory of the ideal form, in order to assess and improve their performance.

The ability to memorize movement patterns and compare them to recent movements is mediated by proprioceptive working memory, a cognitive process that temporarily holds movement information for rapid retrieval and application [1]. While noise in proprioceptive sensors can impair immediate perception of a guided movement, limits in proprioceptive working memory may further compromise the ability to retain and retrieve a desired movement pattern.

Several studies have characterized aspects of proprioceptive working memory and their relevance for skill acquisition. Evidence suggest that memory demands introduce additional error into proprioceptive tasks. For instance, when proprioception was assessed using two common methods—memorizing an experienced limb position versus immediately matching it with the contralateral limb—errors between the tasks were only weakly correlated, with memorized errors being significantly larger [2]. Another similar study also found that memorized limb positions were less accurate than simultaneously reproduced positions [3]. However, in our own work studying a contralateral hand trajectory reproduction task, we found that introducing a memory component did not further impair estimates of hand speed, at least for the first submovement experienced [4].

Proprioceptive memory seems to have limits in how much it can remember, as well. For example, the number of elements presented in a memorized elbow position sequence affected reproduction accuracy. Sequences with > 4-5 elements led to greater errors, likely due to the capacity limits – or saturation – of proprioceptive working memory [5]. When examined as a series of hand movements (pointing, making a fist, etc.), movement memory span was similarly limited to ~4 elements [6].

Proprioceptive memory shows features of a working memory, defined as a cognitive memory process with limited capacity and duration [1]. For example, Sidarta et al. tested participant's ability to remember a test set of passively-guided arm positions [7]. Accuracy declined if a motor learning task (reach to hidden target with binary hit/miss feedback) was

C Callaghan is with the Department of Mechanical and Aerospace Engineering, University of California Irvine, Irvine, CA USA.; callaghc@uci.edu

D Reinkensmeyer is with the Departments of Mechanical and Aerospace Engineering and Anatomy and Neurobiology, University of California Irvine, Irvine, CA USA, dreinken@uci.edu

Research supported by NIH Grant 901HD062744 and NIDILRR Grant 90REGE0010

BioRobotics Lab, Department of Mechanical and Aerospace Engineering, University of California Irvine, Irvine CA 92617

performed between the sample position set and the probe position, indicating the vulnerability of proprioceptive working memory to interference. This study also discovered a positive correlation between the accuracy of proprioceptive working memory and the motor learning rate in the reach-to-hidden-target task, suggesting that having a stronger proprioceptive working memory promotes motor learning.

Building on these findings, a recent study tested the relative importance of proprioceptive memory accuracy versus proprioceptive acuity in motor learning [8]. Participants identified whether a probe arm position was among a test set – an assessment of the accuracy of proprioceptive working memory, and also judged whether one position was more extended than the previous – an assessment of proprioceptive acuity. Motor learning was evaluated by having participants memorize and reproduce a 10-second arm trajectory, with accuracy of the reproduction assessed after being passively guided through the pattern 5 times and then again after experiencing it 35 times. As in Sidarta et al., strong proprioceptive working memory, as measured by the ability to recall previous arm positions, positively correlated with motor learning rate. However, proprioceptive acuity did not correlate with motor learning rate, suggesting that ability to retain proprioceptive information, rather than the accuracy of position sense, is most important for guiding motor learning. Even if movements are sensed accurately, if they cannot be recalled accurately, the movements will not be learned.

These studies show that proprioceptive working memory is error-prone, limited, and vulnerable to interference, while at the same time being important for motor learning. An issue that remains unclear are the specific sources of error introduced by proprioceptive memorization, such as how trajectory shape and timing may be distorted as a function of the duration or complexity of the movement. In our previous study mentioned above in which we found no effect of memorization for position and velocity estimation accuracy, we focused only on analyzing the first submovement of the reproduced trajectory [4]. Here, we performed further analysis of the same experiment, which required participants to use one hand to reproduce a random, template motion applied to the other hand by a servomotor. In addition, we performed a second similar experiment that systematically varied the duration of the template movement.

## II. METHODS

### A. Experiment setting and participants

We conducted the experiments described here during two offerings of an undergraduate engineering laboratory course in two sequential years. The educational purposes of the experiments were to teach concepts and methods in mechatronics, feedback control, data processing, and statistical analysis. We did this in the context of measuring the performance of the human sensory motor system at a proprioceptive task. These experiments qualified for UC Irvine's Self-Determined Exemption as 'Research conducted in educational setting involving normal educational practice.' 212 students participated in Spring 2021, when it was taught remotely, and 33 participated in an in-person offering in Spring 2024. For the remote offering of the course, we mailed students living in the United States a parts kit. Seven international students could not have a kit mailed to them and instead purchased parts locally; their data was omitted. Students were informed they could request their data be excluded from the research study; two students did so. 12 submitted incomplete data. The remaining 191 data sets were analyzed. For the in-person cohort, identical set-ups were assembled by teaching assistants. No participants in the in-person cohort requested their data be omitted and all completed the experiment, resulting in 33 data sets.

### B. Experimental protocol – Experiment 1

For the remote cohort, participants assembled a test apparatus from a parts kit and an instruction manual (Fig. 1A). A teaching assistant verified their set-up before beginning the experiment. A servomotor with position feedback (FEETECH Standard Servo FS5103B-FB) controlled by a microcontroller (Arduino UNO R3) rotated in five-second templates composed of the sum of four sine waves of randomly chosen amplitude, frequency, and phase shift (Fig. 1B). A library of 80 templates were created. Participants attempted to reproduce the template movement by turning the knob of a rotational potentiometer with their other hand (Comimark WH148 Type B10K) (Fig. 1A).

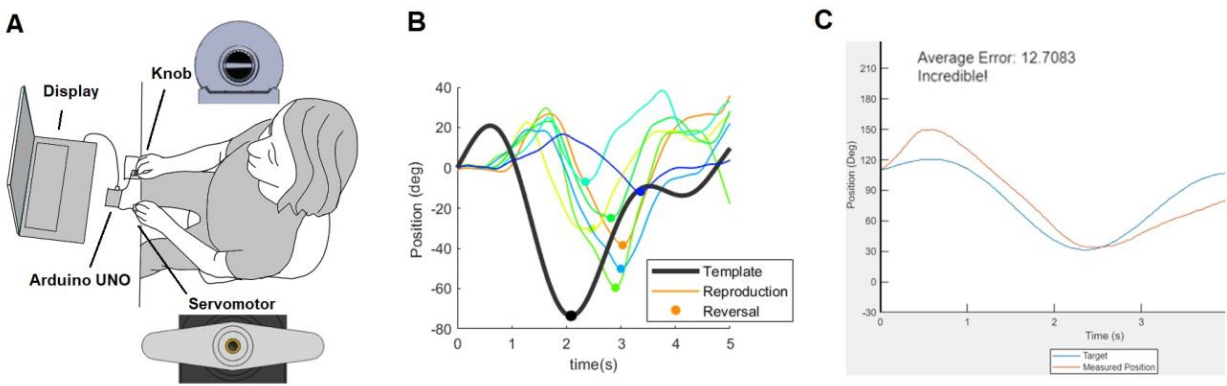

Figure. 1. Overview of experimental apparatus and data analysis, reproduced from [4]. A: Apparatus consisted of an Arduino UNO microcontroller, a servomotor which rotates through a randomized sum of sine waves 2-8 seconds long (a template pattern), a potentiometer knob which participants rotate to reproduce that pattern, and a computer which displays the Matlab experiment interface. B: Example of a template trajectory and simultaneous reproduction (SimRep) attempts. The template's first reversal > 30 deg from the start position, and the associated peak in the reproduction, were used as the 'initial peak' for comparison purposes. Both were identified using Matlab's 'findpeaks' function. C. Example of feedback provided to participants. Every third run the previous three runs' results were displayed as the template and reproduction trajectories, the mean position gap across the duration, and a rating to motivate participants to match well (here, 12.7 deg error was 'incredible'). Three such plots were shown for 10 seconds before proceeding to the next template reproduction run.

Participants were randomly assigned into two groups. One group had a primary condition of simultaneous reproduction (SimRep) and the other had a primary condition of memorized movement reproduction (MemRep). In the SimRep condition, they moved simultaneously with the servomotor. In the MemRep condition, they rotated the potentiometer immediately following completion of servomotor's template movement.

For four sequential days they performed 12 reproduction trials in their primary condition, along with 12 trials in the other condition (i.e. MemRep condition if they were assigned to the SimRep group and vice versa) on the first and fourth days. They experienced a new template each trial; that is, no participant experienced the same template twice in the same condition.

Matlab generated the visual interface which guided participants through the procedure; this interface communicated with Arduino IDE via Serial. The Matlab interface also recorded position data from the servo and potentiometer knob at 60 Hz, which was stored in local files that participants turned in.

Each reproduction began with the servomotor moving to the initial angle for the template; participants matched the potentiometer angle to this position with the assistance of cues to rotate clockwise or counter-clockwise until positions matched within 3 deg. This was the method for participants to indicate they were ready for the next template to begin. If the next template's initial position was within 3 deg of the potentiometer's final position from the previous trial, such that the template would begin without any intervention from the participant, that template was swapped for another. As a result, not all templates were viewed by all participants, nor did all templates have an equal number of reproduction attempts. Each template was experienced by a mean (SD) of 95 (50) participants. Only templates which were experienced by 12 or more participants were included; 60 templates met this criteria.

Once the potentiometer and servo were aligned, the visual interface displayed a 3 s countdown for template movement onset. One template consisted of 5 s of servomotor movement; for SimRep, participants rotated the potentiometer to match this pattern while the servomotor was rotating. For MemRep, participants experienced the entire template first while holding the potentiometer at the template's initial position. MemRep reproduction attempts were considered 'initiated' once the potentiometer had been rotated > 3 deg from its start position. After a run ended, there was a short countdown (3 s) before the servo moved to the start position of the next template. After three runs, the trajectories of the last three attempts were displayed for 10 seconds (Fig. 1C) as well as a rating message based on performance to motivate participants to track well

*C. Experimental protocol – Experiment 2*

In the second experiment, we created 12 unique templates using the same technique as in Experiment 1, but each 8 seconds long (Fig. 2). A teaching assistant assembled the test apparatus, explained the experiment protocol, and supervised data collection which occurred on a single day. Each participant performed 12 trials in the SimRep condition and 12 in the MemRep condition, with the order of condition randomized. For each trial, the duration of the template displayed was randomly chosen from the set {2, 4, 6, 8}, such that each student experienced three templates of each duration.

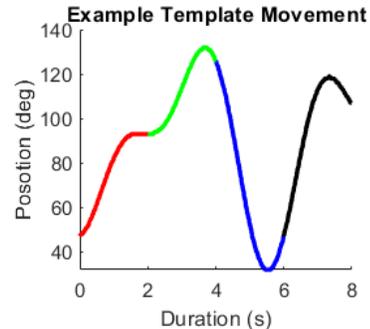

Figure. 2. Exemplar templates of 2, 4, 6, and 8 second durations. The 2-second duration template was identical to the first 2 seconds of the 4, 6, and 8 second duration template; the 4-second template was identical to the first 4 seconds of the 6 and 8 second templates, and so on.

*D. Data analysis*

Position error was quantified as the mean angular difference between the template and reproduced trajectories. For SimRep, both signals were recorded simultaneously. As template and reproduction signals were recorded consecutively under MemRep, prior to analysis, the signals were time-aligned at the moment the servomotor started (for the template) and the moment the potentiometer rotated > 3 deg from start position (for reproduction).

To assess proprioception errors related to the interaction of template characteristics and template duration, templates were segmented between direction reversals to create sets of submovements. For some analyses, we focused on the first submovement, defined as the portion between movement onset and the first reversal > 30 deg from the initial position (Fig. 1B).

We defined an element omission as having occurred if there were fewer direction reversals in the reproduction than in the template. We defined a directional error as having occurred if a participant initially rotated the potentiometer at least 10 deg in the incorrect direction. We quantified timing error as how early (lead) or late (lag) the reproduction's first velocity reversal occurred compared to the template. We defined movement cessation as having occurred if the velocity dropped below 3 deg/s for the remainder of the reproduction signal with > 1 s of reproduction time remaining.

Correlation between template and reproduction signals was quantified for a sliding 1-second window, in step sizes of 0.10 seconds, in both the velocity and position domains. The improvement or decay of correlation over time was quantified with Spearman's correlation (alpha = 0.05), as was the evolution of timing errors for SimRep.

The mean error across the duration of the trajectory and timing errors for the MemRep and SimRep conditions were compared through two-tailed t-tests. Duration effects were quantified through ANOVA, with reproductions categorized by their duration. Relationships between template characteristics and errors were compared using ANCOVA of linear fits.

## III. RESULTS

Participants (N = 191 in Experiment 1, N = 33 in Experiment 2) used their dominant hand to try to reproduce rotational movements displayed to their non-dominant hand by a servomotor, either simultaneously with the template movement (SimRep) or immediately after (MemRep). The template movements followed randomized patterns composed by adding sinusoids; in Experiment 1, template duration was 5 s, while in Experiment 2 template duration varied between 2-8 s. We analyzed how the addition of the memory requirement affected reproduction errors.

### A. Error was significantly higher for MemRep than for SimRep, and increased over time.

Participants produced significantly greater average error in MemRep than SimRep in both Experiment 1 (Fig. 3 A $p < 0.001$), 29.3+26.3 deg versus 21.5+19.4 deg) and Experiment 2 (33.8+29.5 deg vs 23.6+20.5 deg), and SimRep and MemRep error were significantly correlated (Fig. 3B). Error increased over time for MemRep reproductions (Fig 4A, B), but errors under SimRep condition stabilized after 2 seconds. Additionally, reproductions of longer templates in the MemRep condition had significantly higher error than reproductions of short templates (Fig 4B).

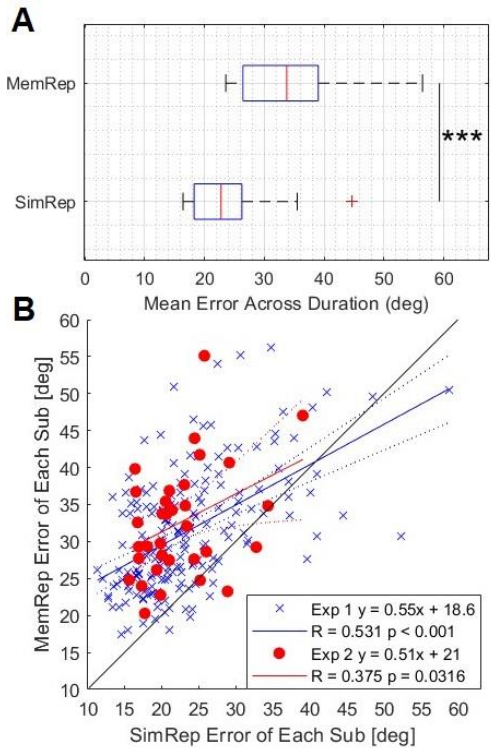

Figure 3. Comparison of SimRep and MemRep mean error. A: In experiment 1, MemRep produced significantly higher error than SimRep (33.8 deg vs 23.6 deg). B: For both experiments 1 and 2, MemRep and SimRep error were significantly correlated.

To identify how long a template must be before there was a significant difference between SimRep and MemRep performance, we broke the templates into 0-2 sec, 2-4 sec, 4-6 sec, and 6-8 sec segments and compared participants' performance for segments. MemRep condition produced significantly higher error for all segments (two-tailed t-test, all comparisons returned $p < 0.001$, Fig. 4B).

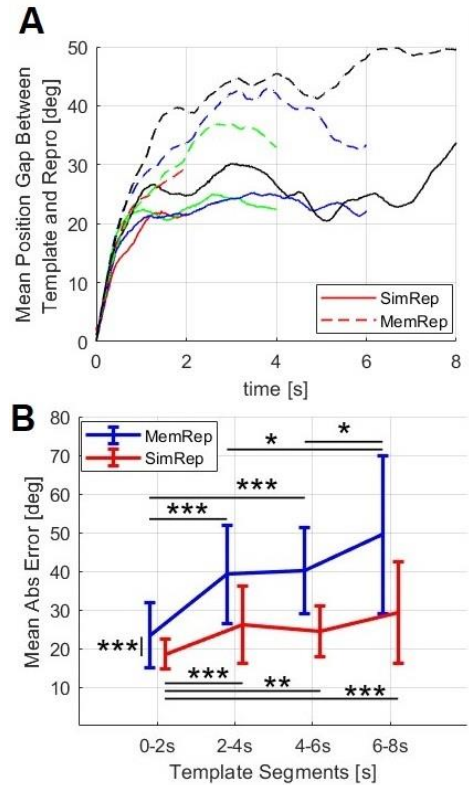

Figure 4. Evolution of error over the duration of the reproduced trajectory, for Experiment 2. A. Mean absolute error at each sample, separated according to the duration of the template being reproduced. B. Mean error for 0-2s, 2-4s, 4-6s, and 6-8s segments, distinguishable in the figure by how long they last. As can be seen in both graphs, error in the SimRep condition stabilized after 2 seconds, while error in the MemRep condition continued to increase over time. Even in the first 0-2 seconds, the MemRep condition produced significantly higher error.

### B. Correlation of template and MemRep reproductions decayed within 3-4 seconds.

Another way to assess the quality of the reproduction is to quantify the strength of its correlations with the template. Under the MemRep condition, correlations between 1 s moving window segments of the template and reproduction velocity trajectories fell below 0.2 within ~2 seconds (Fig 5A).

In contrast, in the SimRep condition, participants improved their correlation as the movement progressed, becoming better at reproducing the displayed trajectory as it occurred. The mean correlation coefficient between SimRep templates and reproduction velocities increased as an exponential rise ($R^2 = 0.75$) that settled at $R = 0.55$ by 3.0 seconds (Fig 5B).

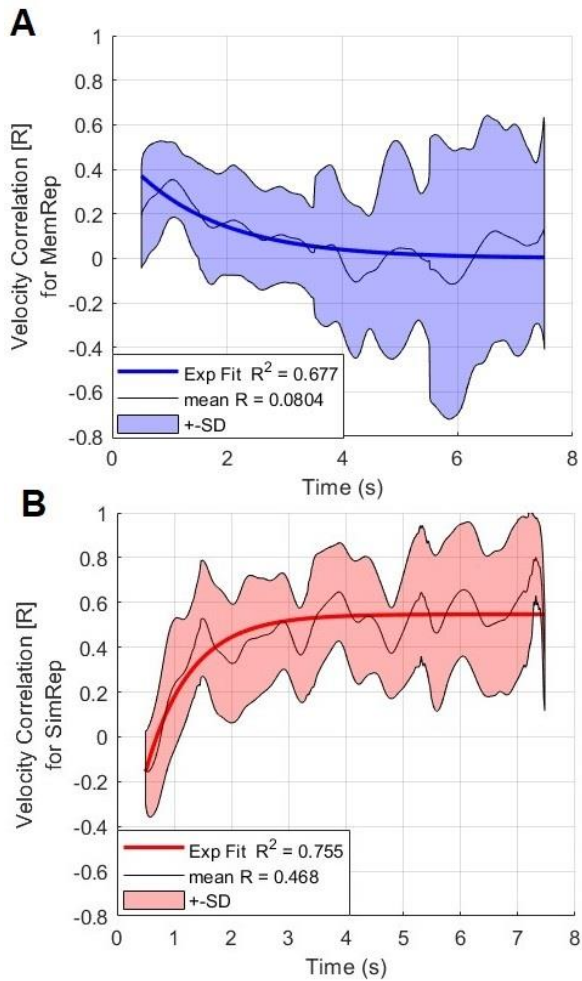
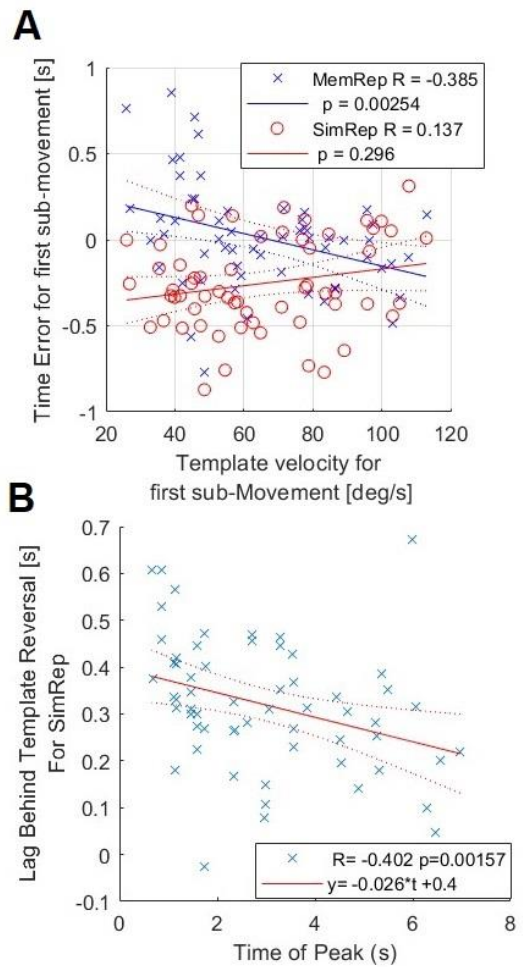

Figure 5. Correlation of template and reproduction velocity for Experiment 2. A. For the MemRep condition, the correlation strength degraded quickly, falling below R = 0.2 within 3-4. Decay followed a first-order exponential decay trajectory. B. For the SimRep Condition, correlation initially increased, then plateaued.

Figure 6 Lead trends differed for MemRep and SimRep A. In experiment 1, in MemRep, participants lead slow movements (positive timing error) and lagged behind fast ones (negative timing error). By contrast, under SimRep, reproductions lagged similarly regardless of the n = 60 templates' speeds. B. In experiment 2, In the SimRep condition, reproductions lag behind the template but magnitude of the lag significantly decreased over time.

### C. Memorization added temporal distortions

Examining the first submovement in the MemRep condition, participants reproductions led slower submovements in the template, peaking before the templates did, and lagged faster ones, peaking after they did (Fig. 6A). In contrast, under SimRep, the first submovement always lagged. The lag was not related to template velocity, averaging -0.27 s + 0.27 (Fig. 6A); and lag decreased significantly over time (6B).

### D. Direction errors were an additional source of error for memorized reproduction

Directional errors were more common under MemRep condition than SimRep, especially if the initial movement was small (Fig. 13). For the SimRep condition, participants rarely made initial direction errors (3.2% of all reproductions); that is, they generally moved in the same direction as the servomotor when reproducing the signal simultaneously. However, for the MemRep condition, participants initially moved in the wrong direction for 9.5% (+/- 13%) of reproductions, a significantly greater rate (p = 0.004). They did this more often when the initial movement was small in magnitude (Fig. 7). For the smallest initial movements (<30 deg), the rate of direction error was ~13%. Reproduction trajectories in the SimRep condition showed no significant correlations between direction error rate and template submovement magnitude or velocity.

We suspected that these direction errors may have contributed to the template and reproductions decoupling over time, but omitting reproductions with direction errors had little effect on the mean correlation across the entire template (R = -0.437 vs R = -0.438, p < 0.001 for both).

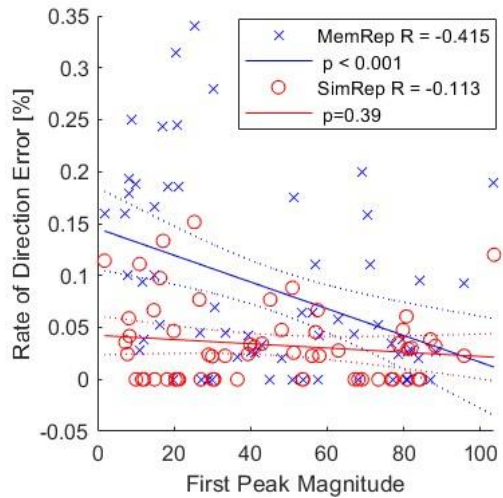

Figure 7 Rate of direction errors for the first submovement. Initial submovements which were small in magnitude were more likely to be misremembered, here quantified as rate participants initially moved the wrong direction for at least 10 deg for each of the n = 60 templates.

### E. Participants stopped moving for longer templates during memorized reproduction

We observed premature movement cessation in the reproductions in the MemRep condition for movements of 4 s or longer (Fig 8A shows several examples), with the rate of movement cessation increasing with the template duration (Fig. 8B). For the longer 6 and 8 s templates, individuals stopped moving ~2 s before the end of the template (Fig. 8B). In the SimRep condition, cessation was not detected; participants continued to move for the duration of the template movement.

### F. Ceiling on submovement reproduction

We characterized the trajectories that were haptically displayed by the number of submovements they had, which varied from 1 to 6. We then compared the number of submovements in the reproduced trajectory to the number in the template trajectory. In the MemRep condition, participants omitted at least one submovement 37% of the time, with the rate of omission increasing with the number of submovements

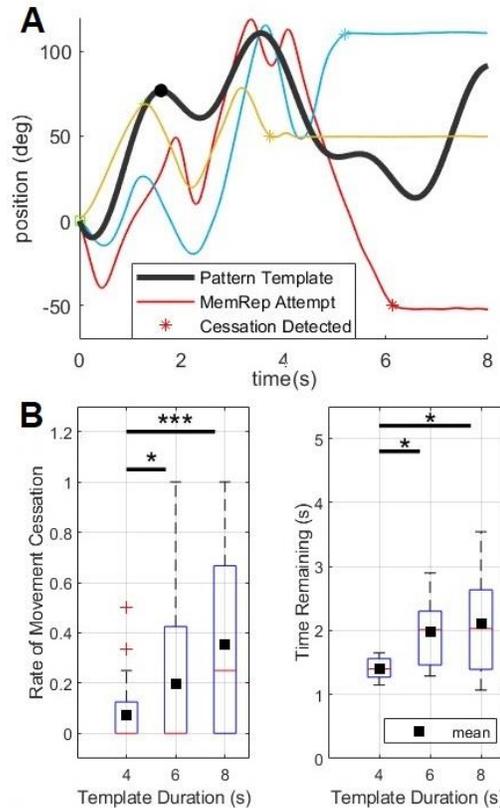

Figure 8. Premature movement cessation. A) Example of reproductions for which the participant stopped moving before the end of the template. B) Rate of direction errors for the first submovement, and time remaining in the template. Initial submovements which were small in magnitude were more likely to be misremembered, here quantified as rate participants initially moved the wrong direction for at least 10 deg for each of the n = 60 templates.

in the template (Fig. 9A), reaching 80% for templates of 6 reversals. They accurately matched the number of submovements in templates with 4 or fewer submovements (Fig. 9B). However, they produced reproductions with an average of 4 submovements for templates with 5 or more submovements, indicating a ceiling on submovement reproduction. Returning to the premature cessation phenomenon described in the previous section, rate of cessation was significantly higher for templates of 5 or more peaks (14.0

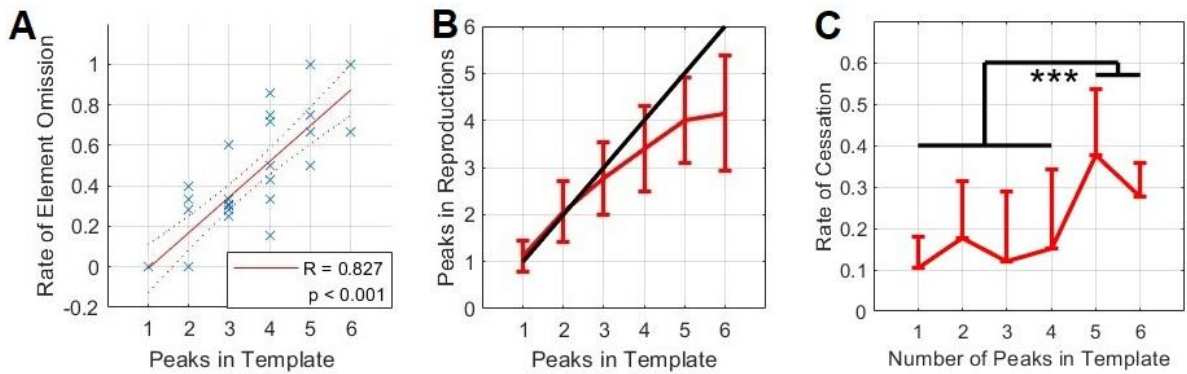

Figure 9. Rate of direction errors for the first submovement. Initial submovements which were small in magnitude were more likely to be misremembered, here quantified as rate participants initially moved the wrong direction for at least 10 deg for each of the n = 60 templates.

+- 15%, vs 35.4 +- 14.7%, p < 0.001, Fig. 9C), providing further evidence of memory capacity limitation related to number of submovements.

## IV. DISCUSSION

The goal of this study was to identify sources of proprioceptive error introduced by including a memorization component into a contralateral hand movement reproduction task. Memorized reproductions were significantly less accurate than simultaneous reproductions, even in the first 2 seconds, and accuracy decayed rapidly during memorized reproduction. We identified several specific sources of error that affected only memorized reproductions: systematic distortions in timing, misrecalls of initial submovements (especially smaller ones), and a maximum capacity of 4 submovements. We discuss these results now, followed by implications and future directions

### A. Incorporating a memorization task introduced timing errors, direction errors, element omissions, , and movement cessation.

Memorized reproductions showed systematic temporal distortions: participants led slow movements and lagged fast ones. In a companion paper (paper in submission to IEEE Transactions on Haptics, available as pre-print [4]) that examined some of the same data, participants overestimated the speed of the slowest rotations by ~45% and underestimated the speed of the fastest by ~30% during simultaneous reproduction. These velocity-estimation biases carried over to but were not amplified by memorization. Thus, the additional timing error observed here for memorized compared to simultaneous reproduction cannot be explained by velocity misestimation alone. Instead, participants tended to move too fast and too long on slow segments, and too slowly and too briefly on fast segments.

Direction errors occurred at a higher rate with smaller magnitude template movements, demonstrating that subtle movements are more likely to be misremembered. While large movements are typically associated with higher error for both simultaneous [3], [9] and memorized [3] joint position reproduction tasks for single-element movements, when tasked to memorize a movement sequence, we found that subtle movements were often misremembered but were rarely misinterpreted during SimRep.

Direction errors were associated with the omission of the first movement. This may have caused subsequent submovements to begin 'early'. However, even after omitting reproductions with direction errors, there was a trend of leading slow movements and lagging fast movements. This suggests an innate reversal rate participants gravitated toward, rather than accurately replicating template reversal timing. Temporal errors affected subsequent movements by also shifting them early or late, leading to temporal divergence between template and reproduction.

Movement cessation occurred significantly more often for templates of > 4 reversals, or > 4 seconds, indicating that proprioceptive working memory is limited and quickly becomes saturated. The element omission analysis also indicated a limit to proprioceptive working memory of ~4 direction reversals, which is equivalent to transitions between 5 submovements. Horváth et.al also found 5 elements to be the limit of proprioceptive memory before accuracy decays, though that experiment involved memorization of a position sequence [10]. Smyth and Pendleton similarly found that recall of a hand task sequence was limited to four tasks [6]. We did not assess whether element accuracy decayed over time, only whether the correct number of elements could be recalled and whether elements were omitted from the start or the end of the sequence. Memory span was limited for late-sequence elements in that elements after the 4th reversal were forgotten, and the remainder of the sequence was omitted. Early-sequence movements were vulnerable to being forgotten if they were small in magnitude. From this findings, sequence recall is vulnerable for both early and late elements, but those vulnerabilities differ in cause: late elements are lost to memory saturation, and subtle movements are lost in favor of large ones.

That four reversals was also the memory span limit for our experiment suggests the template movement was segmented in participant memory, with reversals indicating the transition between segments. Future research that applies template memorization to a 2-dimensional space, where trajectories may reverse in one axis while continuing uninterrupted in another, would provide insight into movement trajectories are encoded in working memory.

### B. SimRep's plateaued correlations suggest participants use haptic feedback to anticipate template behavior

In the SimRep condition, we observed a surprising result: people got better as the reproduction progressed. Correlation initially improved then settled within ~3 seconds. This is also when error stabilized; the initial movement generated significantly more error than subsequent movements. One possible explanation is that participants act on haptic feedback to anticipate template behavior, a source of information not available for the first submovement.

Though both MemRep and SimRep conditions produced timing errors, under SimRep, timing errors consisted of lags and these lags decreased over time. We attribute lagging to reaction time; processing the haptically experienced movement of the servomotor and reacting to it is not an instantaneous process. Further, when a template begins the participant has no indication whether the servomotor will begin with clockwise or counterclockwise movement. When movement begins they must interpret not only the initial velocity but the initial direction of movement before the movement can be transferred to the opposite hand. Though direction errors for SimRep were rare, the time to perform this processing is non-trivial. Due to the sinusoidal nature of the templates used, when the servomotor approaches a reversal it will begin to slow (exhibiting negative acceleration). This may be providing haptic information to the participant to anticipate the template direction is about to change. As the servomotor could only switch between clockwise and counterclockwise, this deceleration also primes the participant to move in a specific direction, which would not have been the case in a two-DOF movement sequence.

*C. Memorized movements were poorly correlated with their templates*

The errors unique to MemRep induced rapid decay of correlation between a template and its reproductions. Direction errors and timing errors shifted the template and reproductions out-of-phase, generating mixed positive and negative correlations which averaged to no correlation (R ~ 0). Without access to haptic feedback on template speed or reversal timing, participants had no information to strategize compensating for these errors, allowing error to cascade. Templates with more than four reversals had high rates of movement cessation, leaving long periods of no movement (and therefore R = 0). From the first movements to the last, correlation between MemRep template and reproduction started poor and decayed to no correlation within seconds.

*D. Implications and future directions*

In the Introduction, we observed that guided movement training approaches (such as those used in rehabilitation, dance, and sports training) rely on accurate encoding and recall of proprioceptively demonstrated movements. Our findings suggest three practical implications. First, expect early reproductions to show systematic temporal distortions – slow segments are sped up and fast ones slowed down. Understanding this may be useful for targeted correction. Second, subtle, low-amplitude movements are more difficult to learn; they likely require increased attentional focus and/or isolated practice. Third, in the initial stages of training, limit demonstrations to short, simple chunks ≤ 4 elements and 3-4 seconds in duration. Individuals can piece together longer movement sequences, as is evident in dance performances. But these longer routines might best be built by chaining stabilized elements, improving accuracy before complexity.

These recommendations assume that effects observed for relatively simple hand movements generalize to more complex, functional movements. In real-world movement training, guided movements are not arbitrary, they're useful. Extending the types of analysis used here to task templates that mimic relevant activities (e.g. transporting a spoon from bowl to mouth) can likely yield further actionable data on the limitations of demonstration-and-replay training. In this context, the error sources identified here – timing mismatch, omission of movement elements, and premature cessation – offer practical categories for monitoring performance, evaluating progress, and tailoring feedback.